\documentclass[useAMS,usenatbib]{mn2e}
\usepackage{amsfonts}
\usepackage{amssymb}
\usepackage{graphicx}
\usepackage{color}
\usepackage{ulem}
\def\tablenotemark#1{\rlap{$^{\rm #1}$}}
\newcommand\phn{\phantom{0}}%
\newcommand\phd{\phantom{.}}%
\newcommand{\myemail}{xjjiang@nju.edu.cn}

\newcommand{\Lsun}{\ensuremath{L_\mathrm{\odot}}}
\newcommand{\LIR}{\ensuremath{L_\mathrm{IR}}}
\newcommand{\hcop}{HCO\ensuremath{^+}}

\newcommand{\kms}{$~\rm km~s^{-1}$}
\newcommand{\apj}{ApJ}
\newcommand{\apjl}{ApJ}
\newcommand{\mnras}{MNRAS}
\newcommand{\aj}{AJ}
\newcommand{\aap}{A\&A}
\newcommand{\apjs}{ApJS}
\newcommand{\araa}{ARA\&A}

\title{Dense Molecular Gas in Nearby Gas-rich Active Galaxies}

\author[X. Jiang et al.]{Xuejian Jiang\thanks{E-mail: \myemail},
Junzhi Wang, Qiusheng Gu\\
$^1$ Department of Astronomy, Nanjing University, Nanjing 210093, P. R. China \\
$^2$ Key Laboratory of Modern Astronomy and Astrophysics, Nanjing University, Ministry of Education, Nanjing 210093, P. R. China}

\begin{document}

\date{Accepted 2011 August 7.  Received 2011 August 4; in original form 2011 March 24}

\pagerange{\pageref{firstpage}--\pageref{lastpage}} \pubyear{2011}

\maketitle

\label{firstpage}

\begin{abstract}
We present 3 mm observations obtained with the IRAM 30-m telescope for ten nearby gas-rich active galaxies spanning three orders of magnitude in infrared luminosity. Emission lines of HCO$^+$(1-0), HCN(1-0), and C$_2$H(1-0) were simultaneously detected in most galaxies of our sample. 
We also tentatively detected the rare isotopic species, H$^{13}$CN, HC$^{15}$N and H$^{13}$CO$^+$ in up to four galaxies (IC\,694, NGC\,3690, NGC\,4258 and NGC\,6240). Our estimation shows that the optical depth of HCN is low to moderate in these galaxies  ($\tau \sim 1-5)$. After comparing the intensity ratios of different molecular emission lines (HCN/HCO$^+$, C$_2$H/HCN, and C$_2$H/\hcop) with the infrared luminosity (\LIR), we find that in the infrared luminous sample of five composite (AGN+Starburst) galaxies, these ratios vary with \LIR: the HCN/HCO$^+$ ratio increases with \LIR, which is consistent with previous studies, while there is a slight trend in the composite galaxies that both C$_2$H/HCN and C$_2$H/HCO$^+$ ratios decrease with increasing \LIR. Although our sample is limited, this trend may possibly imply a relation between the overall gas density and the infrared luminosity of galaxies.
\end{abstract}


\section{Introduction}
Dense molecular gas plays an essential role in the formation and evolution of galaxies.  It could supply the material for starbursts (hereafter SB) and for accretion in super massive black holes (active galactic nuclei, AGN) in galaxies. It is also critical in understanding galaxy evolutionary processes such as galaxy interaction, AGN feedback, etc. In return, its properties are strongly affected by these phenomena and their accompanying radiation fields, such as X-ray or UV radiation. Recently, both observational and theoretical studies have discussed the relation between the environment or activity of host galaxies and the physical or chemical properties  of molecular gas \citep{henkel91, krumholz07, krips08, baan08}.

Recent studies have developed some molecular gas tracers, based on millimeter observations, to investigate the intrinsic properties of active galaxies. One advantage of these molecular tracers is that their emission lines are less affected by dust extinction than optical lines. This is especially important for the study of (ultra) luminous infrared galaxies ((U)LIRGs), where  substantial amount of dense gas and dust are affected by ongoing violent star formation and/or a powerful AGN. 

While CO(J=1-0) can trace the general molecular hydrogen gas content of low to medium density, molecules with high dipole moment only trace much denser molecular gas. Nowadays, dense gas tracers, such as HCN, HCO$^+$, and CS, have been observed and analysed extensively for the study of the densest molecular gas content ($n(\rm H_2)>10^5~\rm cm^{-3}$) in galaxies \citep[e.g.,][]{gao04a, gc08, baan08, bayet09}. A tight linear correlation between the infrared luminosity (\LIR) and HCN(J=1-0) line luminosity ($L'_{\rm HCN(1-0)}$) has been found, indicating an intimate relation between star formation rate (SFR) and the dense gas mass \citep{gao04b}. This correlation has been verified over eight orders of magnitude of infrared luminosity,  from the molecular clumps in the Galaxy to LIRGs \citep{wu05}, and from nearby galaxies to high-$z$ galaxies and QSOs as well \citep{gao07}. On the other hand, it has been proposed that the chemical and/or physical properties of these molecules vary with the types of galactic activity. For instance, HCN/CO(J=1-0) and HCN/HCO$^+$(J=1-0) intensity ratios were both found to be higher in AGN-dominated systems compared to SB-dominated ones \citep[e.g.,][]{kohno05,krips08}. This may be due to X-ray dominated regions (XDR) produced by AGN which affects the abundance of HCN and/or HCO$^+$ \citep{gc06}. However, it has also been demonstrated that there is a wide range of excitation among different galaxies, implying that only adopting HCN(J=1-0) or HCO$^+$(J=1-0) as dense gas tracers could be insufficient. Thus there is an urgent need for data of additional  molecular species and transitions, to constrain the relation between dense gas, star formation and AGN activities \citep{greve09}.  Most recently, extensive surveys have been focusing on other molecules, such as HC$_3$N, providing additional diagnostic tools for the study of abundance and excitation properties of galaxies \citep{costagliola11, lindberg11}.

In this work we present IRAM 30m single-dish observation towards ten nearby active galaxies. HCN(J=1-0), HCO$^+$(J=1-0), and C$_2$H(N=1-0) emission lines were simultaneously detected in most sources.  
With the simultaneous detections and measurements of these molecular lines, we compare the intensities of HCN, HCO$^+$, and C$_2$H for various galaxies and discuss the relation between C$_2$H line luminosity and infrared luminosity of galaxies.

Throughout this paper, we adopt a flat cosmological model with $\Omega_{\rm M} = 0.27$ and $\Omega_{\Lambda} = 0.73$, and a Hubble constant $H_0$ = 73\ km s$^{-1}$ Mpc$^{-1}$ \citep{bennett11}.

\section{Observations and data reduction}


We selected ten gas-rich active galaxies (see Table \ref{tbl-source}) to  observe molecular lines at a rest frequency around 87 GHz in July 2009 with  the IRAM 30m millimeter  telescope at Pico Veleta, Spain. 
The EMIR receiver with dual-polarization, the WILMA back-end with 2 MHz  channel spacing ( this corresponds to  $\sim6.9$ km/s for lines near 87 GHz) and 3.7 GHz frequency coverage  were used for this observation. The measurements were carried out using the standard wobbler switching mode with beam throws of $\pm120''$ and a switching frequency of 0.5 Hz.  Pointing and focus were checked about every 2 hours by measuring nearby QSOs with strong millimeter continuum emission. The typical system temperature at 87 GHz was less than 100K, including $\sim$ 40K from the receiver. We read out one spectrum every 12 minuets, which gave an effective  on-source integration time of about five minutes.

\begin{table*}
 \centering
 \begin{minipage}{150mm}
  \caption{The Basic Properties of Our Sample Galaxies$^*$ \label{tbl-source}}
  \begin{tabular}{@{}lccrcccccc@{}}
\hline
 Source   & \multicolumn{2}{c}{RA  (J2000)   Dec} &  $V_{\rm Helio}$ &  On-Source time\tablenotemark{a}
           &  RMS\tablenotemark{b}  &  \LIR\tablenotemark{c}&  Type\tablenotemark{d} \\
           &  h\phn m\phn s  & \phn \degr~\phn \arcmin ~\phn \arcsec  &
(km s$^{-1}$) & (min) & (mK) & ($10^{10}$\Lsun)\\
\hline
IC 694   & 11 28 34.2 & $+$58 33 48 & 3100  & 152.5 & 1.1 & 32\tablenotemark{1} &AGN+SB&\\
NGC 3690 & 11 28 31.0 & $+$58 33 48 & 3180  & 111\phn\phd & 1.4 & 25\tablenotemark{1}  &AGN+SB& \\
NGC 6240 & 16 52 58.9 & $+$02 24 04 & 7368  & 225.5 & 1.0 & 61\tablenotemark{1}  & AGN+SB&\\
Mrk  231 & 12 56 14.2 & $+$56 52 25 & 12500 & 133\phn\phd  & 1.3  & 293\tablenotemark{1}\phn  &AGN+SB&\\
Mrk  273 & 13 44 42.1 & $+$55 53 13 & 11315 & 72\phd & 2.0 & 125\phn &AGN+SB&\\
NGC 4258 & 12 18 57.5 & $+$47 18 14 & 500   & 165.5 & 1.0  & \phn\phn\phd1.2\tablenotemark{2}  &AGN& \\
NGC 4388 & 12 25 46.7 & $+$12 39 44 & 2560  & 171.5 & 1.0  & \phn\phn\phd6.1 &AGN& \\
NGC 4593 & 12 39 39.4 & $-$05 20 39 & 2389  & \phn52.5 & 1.8  & \phn\phn\phd2.1 &AGN& \\
NGC 5506 & 14 13 14.8 & $-$03 12 27 & 2000  & 62\phd & 1.7  & \phn\phn\phd2.9 &AGN& \\
NGC 5548 & 14 17 59.5 & $+$25 08 12 & 5120  & 62\phd & 1.3  & \phn\phn\phd3.9 &AGN& \\
\hline
\end{tabular}

\medskip
$^*$This table lists our sample.\\
$^{a}${On-Source observation time.}\\
$^{b}${Baseline RMS (corrected to $T_{\rm mb}$) of the raw spectrum.}\\
$^{c}${Infrared luminosity \LIR~ from 1) \citet{gc06} 2) \citet{rice88} and others calculated from IRAS infrared flux.}\\
$^{d}${Classification of the galaxies.}\\
\end{minipage}

\end{table*}


The data were reduced with the  CLASS program of the  GILDAS\footnote{http://iram.fr/IRAMFR/GILDAS/} package. 
A first-order fitting was used to remove baselines from all combined spectra. The RMS of each spectrum was also obtained when fitting the baselines. The basic physical parameters of our sample galaxies, including original RMS and on-source time, are listed in Table \ref{tbl-source}.

For the spectra, we did a preliminary identification of molecules refering to  frequencies  from the NIST database {\it Recommended Rest Frequencies for Observed Interstellar Molecular Microwave Transitions}\footnote{http://www.nist.gov/pml/data/micro/index.cfm}.

The velocity-integrated intensities of molecular lines are calculated using
\begin{equation}
I = \int _{\Delta V} T_{\rm mb}{\rm d}v,
\end{equation}
where $T_{\rm mb}$ is the main beam brightness temperature, and $\Delta V$ is the  velocity range used to integrate the intensity. Molecular line intensity in antenna temperature is converted to main beam temperature $T_{\rm mb}$ using $T_{\rm mb} = T_a^{*} F_{eff}/B_{eff}$ with forward efficiency $F_{eff}$=95\% and beam efficiency $B_{eff}$=81\% at 86 GHz. The Half-Power BeamWidth (HPBW) in this band was 29$''$. 

\section{Results}
\begin{figure*}
\begin{minipage}{170mm}
\includegraphics[angle=0,scale=.31]{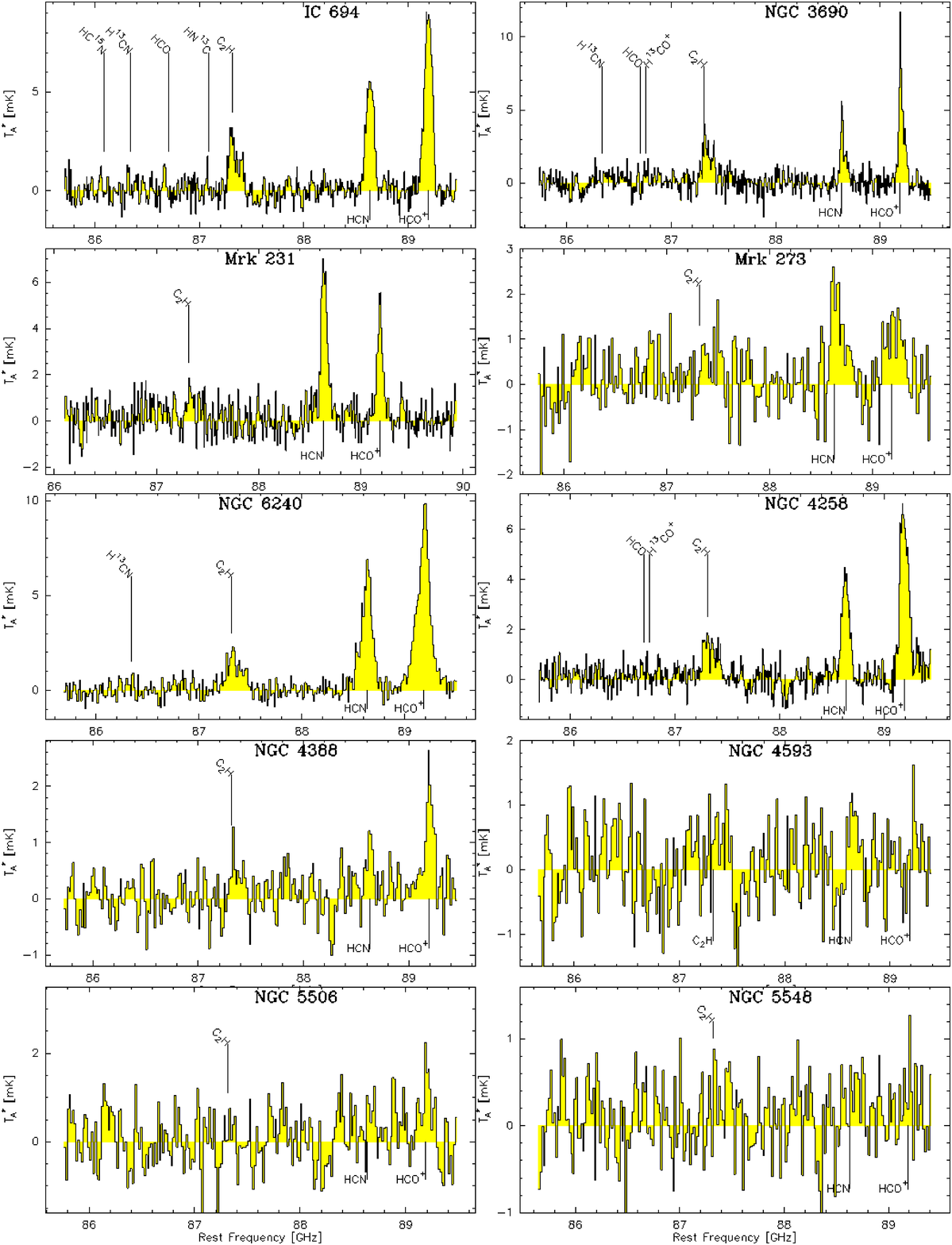}
\caption {Full spectra simultaneously obtained from the IRAM 30m telescope. The bandwidth of the spectra obtained is $\sim$ 4 GHz, and the RMS and spectral resolution are listed in Table \ref{tbl-source}. \label{spec-full}}
\end{minipage}
\end{figure*}

\begin{figure*}
\begin{minipage}{170mm}
\includegraphics[angle=0,scale=.3]{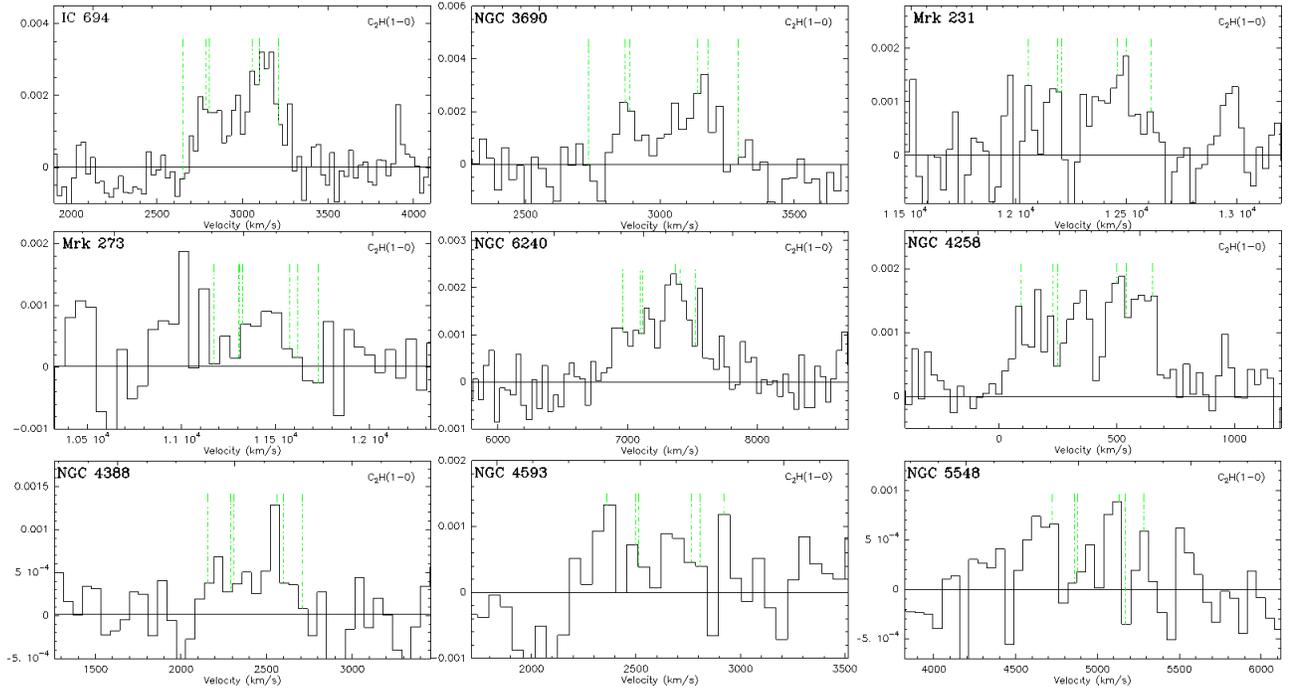}
\caption{Spectra of C$_2$H(1-0) lines. Green dot dashed lines mark the 6 hyperfine frequencies of C$_2$H(1-0). The intensity scale is $T^*_A$ in unit of K.
\label{spec-c2h}}
\end{minipage}
\end{figure*}

\begin{figure}
\includegraphics[angle=0,scale=.9]{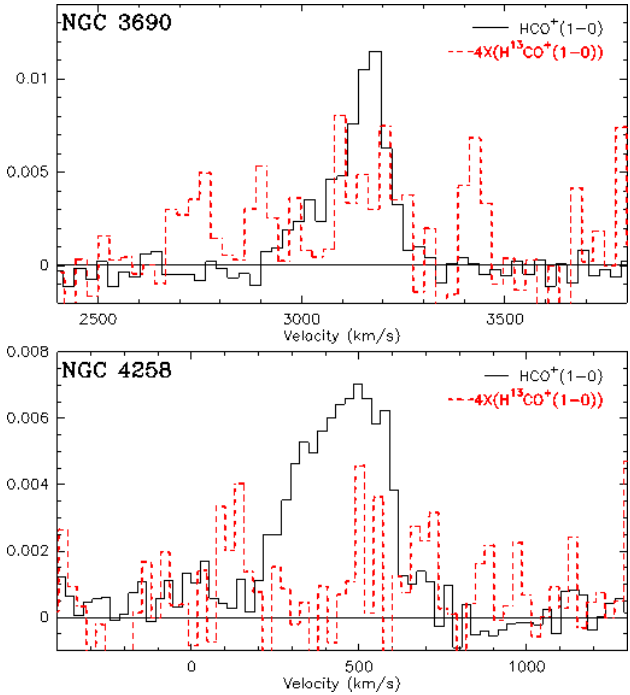}
\caption{Spectra of HCO$^+$(1-0) {\it (black solid lines)} superimposed with the scaled-up isotopes spectra, H$^{13}$CO$^+$(\textit{red dashed lines}) within the same velocity range. The intensity scales is $T^*_A$ in unit of K.
\textit{(See the electronic edition of the Journal for a color version of this figure.)}} 
\label{spec-hcop}
\end{figure}

\begin{figure}
\includegraphics[angle=0,scale=1]{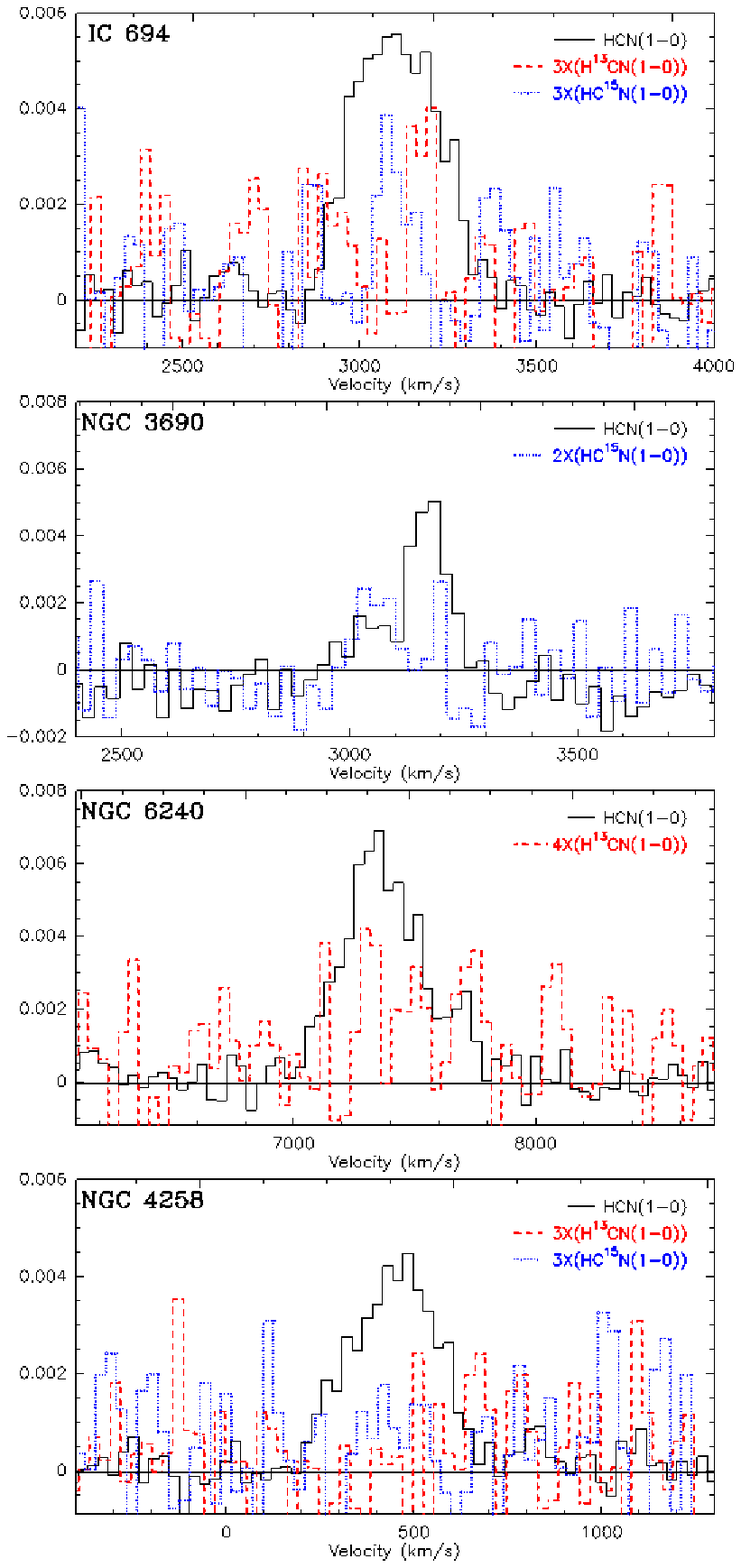}
\caption{Spectra of HCN(1-0) {\it (black solid lines)} superimposed with the scaled-up isotopes spectra, H$^{13}$CN{\it(red dashed lines)} and HC$^{15}$N{\it (blue dotted lines)} within the same velocity range. The intensity scale is $T^*_A$ in unit of K.
\textit{(See the electronic edition of the Journal for a color version of this figure.)}} 
\label{spec-hcn}
\end{figure}

Figure~\ref{spec-full} shows the entire spectrum of each galaxy with the position of certain lines being marked. A more detailed examination   of the spectrum of C$_2$H$(N=1-0)$ lines of each galaxy can be found in Figure~\ref{spec-c2h}. Figure~\ref{spec-hcop} and \ref{spec-hcn} show the spectra of HCO$^+(J=1-0)$ and HCN($J=1-0$) lines along with tentative detections of their isotopes. Table~\ref{tbl-intensity} presents our measurements of the intensities of molecular lines. 

The results of the prominent molecular emission lines HCO$^+$(1-0) and HCN(1-0) for the five (U)LIGRs of our sample are consistent with those observed in \citet{gc08}, and results of NGC\,6240, NGC\,4388 and Mrk\,231 are consistent with \citet{costagliola11}. Meanwhile, we find that some rarely observed species toward extragalactic objects are also detected. At $> 2\sigma$ level, we tentatively detected:  H$^{13}$CO$^+$(1-0) in NGC\,3690 and NGC\,4258, H$^{13}$CN(1-0) in IC\,694 and NGC\,6240, and HC$^{15}$N(1-0) in IC\,694. We present the measurements of these isotopes in Table~\ref{tbl-isotope}. Weak molecular lines are also overplotted upon the strongest line, with a identical velocity range, to be checked if they have a similar velocity distribution. Other molecular lines such as HNCO(5-4) and HCO(1-0) might be detected in some galaxies.

There are two groups of hyperfine features of C$_2$H(1-0), spaced by $\sim$ 300 km/s and having a line intensity ratio of 2:1 in the optical thin case under LTE condition \citep{tucker74}. Although it is impossible to distinguish every single component of C$_2$H(1-0) due to the line broadening of galactic kinematics, the split hyperfine line features can be readily seen in the spectra of most galaxies with prominent C$_2$H detection, and their relative intensity ratio also approximates 2 (See Figure~\ref{spec-c2h}, where the six hyperfine transition frequencies are indicated). However, C$_2$H(1-0) of NGC\,4258 is an exception, the peak intensities of whose split structures are almost equal.

Generally, C$_2$H(1-0) emission line is the third strongest line after HCO$^+$(1-0) and HCN(1-0) within our observation window. Determined from the high S/N spectra, its peak brightness temperature is typically 1/4 to 2/3 that of HCN for our sample galaxies.

Here we present detailed information of some galaxies respectively.

\begin{table*}
 \centering
 \begin{minipage}{150mm}
  \caption{Molecular Lines measurements$^*$\label{tbl-intensity}}
  \begin{tabular}{@{}lccrccccrr@{}}
\hline
Source  & $\delta V$\tablenotemark{a} & RMS\tablenotemark{b}&
$I^{\rm HCO^+}_{1-0}$ & $I^{\rm HCN}_{1-0}$   &  $I^{\rm C_2H}_{1-0}$ &
$\Delta v$\tablenotemark{c} & $\Delta v$\tablenotemark{d} &$V^{\rm HCO^+} $  &$V^{\rm HCN}$\\
  & (km s$^{-1}$) & (mK) &
(K km s$^{-1}$)  &(K km s$^{-1}$) &(K km s$^{-1}$)  &\multicolumn{4}{c}{(km s$^{-1}$)}\\  
\hline
 IC\,694   & 26.892&  0.54 & 3.32$\pm$0.14 &1.94$\pm$0.14 & 1.29$\pm$0.15     &600 &700 &  3096 &   3102 \\
NGC\,3690  & 26.892&  0.72 & 2.15$\pm$0.16 &1.10$\pm$0.16 & 1.20$\pm$0.18     &440 &550 &  3155 &   3165 \\
NGC\,6240  & 40.338&  0.41 & 4.97$\pm$0.18 &2.96$\pm$0.18 & 1.22$\pm$0.16     &1200&900 &  7373 &   7363 \\
Mrk\,231   & 26.892&  0.66 & 1.28$\pm$0.15 &1.81$\pm$0.15 & 0.44$\pm$0.18     &500 &675 & 12510 &  12470 \\
Mrk\,273   & 53.784&  0.70 & 1.02$\pm$0.42 &1.28$\pm$0.42 & $<$ 0.25          &800 &445 & 11290 &  11360 \\
NGC\,4258  & 26.892&  0.49 & 2.39$\pm$0.12 &1.34$\pm$0.12 & 0.78$\pm$0.14     &550 &700 &   450 &    452 \\
NGC\,4388  & 53.784&  0.36 & 0.61$\pm$0.15 &0.26$\pm$0.15 & 0.35$\pm$0.20     &380 &700 &  2534 &   2569 \\
NGC\,4593  & 53.784&  0.63 & 0.19$\pm$0.22 &0.34$\pm$0.22 & $<$ 0.23          &275 &400 &  2394 &   2456 \\
NGC\,5506  & 53.784&  0.61 & 0.44$\pm$0.21 &0.20$\pm$0.21 & $<$ 0.22          &280 &390 &  1972 &  ~1872 \\
NGC\,5548  & 53.784&  0.45 & 0.08$\pm$0.09 & $<$ 0.04     & $<$ 0.06          &100 &200 & ~5100 &  ~5100 \\
\hline
\end{tabular}\\
\medskip

$^*$RMS and molecular line integrated intensity is corrected to $T_{\rm mb}$ using $T_{\rm mb} = T_a^*F_{eff}/B_{eff}$ where $F_{eff}$=95$\%$ and $B_{eff}$=81$\%$ in 86 GHz during our observation. Upper limits are $<$ 2 $\sigma$.\\
$^{a}${Velocity resolution of the smoothed spectrum.}\\
$^{b}${Baseline RMS of the smoothed spectrum at the corresponding velosity resolution. Errors of molecular lines is 1$\sigma$ calculated from $\sigma={\rm RMS} \cdot \sqrt{\Delta v \cdot \delta V}$,}\\
$^{c}${Velocity range used to integrate the line intensities of HCN and HCO$^+$.}\\
$^{d}${Velocity range used to integrate the line intensities of C$_2$H.}\\
\end{minipage}
\end{table*}

\subsection{NGC\,6240}
NGC\,6240 is a local LIRG with extreme starburst activity and the first Binary Active Galactic Nucleus \citep{komossa03}. The detection of H$^{13}$CN line was only marginal and its spectrum indicates complicated velocity components, probably a reflection of the dynamical structure of NGC\,6240 (Figure. \ref{spec-hcn}). Comparing to other galaxies, the HCO$^+$ and HCN lines of NGC\,6240 show very sharp single-peak profile and much broader line width ($>$ 1000\kms). \citet{greve09} reported an extensive molecular line data set of NGC\,6240, including multi-transition CO, HCN, HCO$^+$, and CS. Our observation of C$_2$H(1-0) lines of NGC\,6240 is a complement to their work, because C$_2$H(1-0) has an intermediate critical density lying between HCN and CO and has its own distinctive chemical property.

\subsection{IC\,694 \& NGC\,3690 (Arp\,299)}
Arp\,299, a well-known merging system, is composed of two galaxies: IC\,694 (Arp\,299A) and NGC\,3690 (Arp\,299B+C, $45''$ to the west of Arp\,299A). Here Arp\,299C refers to the overlapping interface of the two merging galaxies, which is also a starburst HII region  \citep[e.g.,][]{gehrz83, aalto97, casoli99}. It is believed that IC\,694 and NGC\,3690 both host an AGN with strong starburst \citep{ballo04,gm06}, although for IC\,694 there is still some controversy \citep{gallais04}. Previous interferometry observations have revealed that the $^{12}$CO(1-0), $^{13}$CO(1-0), and HCN(1-0) emission peak mainly at these three components Arp\,299A, B, and C \citep{aalto97,casoli99}.

The spectrum of IC\,694 shows plenty of molecular lines. Besides the obvious HCO$^+$(1-0), HCN(1-0), and C$_2$H(1-0) lines, H$^{13}$CN(1-0), HC$^{15}$N(1-0), HCO(1-0), and HNCO(4-3) lines were also tentatively detected at $> 2\sigma$ level (see Figure \ref{spec-full}), but H$^{13}$CO$^+$(1-0) was not detected.

Of our observation towards NGC\,3690, the $29''$ beam only yielded to us a spectrum of Arp\,299B and Arp\,299C together. The spectrum shows complicated velocity components. No signal of H$^{13}$CN line appears, and the velocity components of HC$^{15}$N are coincident with those of HCN, even if the spectrum was less smoothed(see Figure \ref{spec-hcn}). However, considering the low S/N  of these two isotopes and the high N/$^{15}$N abundance ratio ($\sim 200 - 400$), these features could be artificial. The H$^{13}$CO$^+$ line might be blended with the hyperfine lines of HCO, as its frequency lies between the four hyperfine structures of HCO. We find that the strongest HCO$^+$(1-0), HCN(1-0), and C$_2$H(1-0) line components of Arp\,299B+C all peak at $V\sim 3180$~\kms, which is corresponding to the starburst region of Arp\,299C. We also find that the HCO$^+$ and C$_2$H lines in the spectrum of NGC\,3690 have higher peak temperatures (T$_R^*$, $\sim 12$~mK for HCO$^+$ and $\sim 4.0$~mK for C$_2$H, respectively) than the corresponding lines in the spectrum of IC\,694 (T$_R^* \sim 9 $~mK for HCO$^+$ and $\sim 3.5$ mK for C$_2$H), but both lines of IC\,694 have higher integrated intensities.

Comparing with previous studies of Arp\,299, our measurement of the flux of HCN line for IC\,694 (9.8 Jy km s$^{-1}$) is slightly lower than that from \citet{casoli99} (13.8 Jy km s$^{-1}$), and our measurement for NGC\,3690 (5.5 Jy km s$^{-1}$) is also lower than that of \citet{solomon92} (8 Jy km~s$^{-1}$) (We use a 5.9 Jy/K conversion factor at this frequency during our observation).
These molecules have been detected in Sgr B2 giant molecular cloud and the star-forming KL region of Orion cloud \citep{turner89}. In their work the peak temperature ratio (T$_R^*$(HCN)/T$_R^*$(H$^{13}$CN)) is $\sim$ 4 for Sgr B2 and $\sim$ 10 for Orion KL, while our results show that this ratio is $\sim$ 4 for IC\,694. And the HCN/C$_2$H ratio is $\sim$ 3.6 in Sgr B2, $\sim$ 6.7 in Orion KL and $\sim$ 1.8 in IC\,694, respectively.

\subsection{Mrk\,231 \& Mrk\,273} 
Mrk\,231 \& Mrk\,273 are local ULIRGs.  It is suggested that $\sim 1/3$  the bolometric luminosity of Mrk\,231 is contributed by AGN, and $\sim 2/3$   by starburst in the molecular ring or disk \citep{downes98,farrah03}. With respect to Mrk\,273, radio observations have revealed an extreme compact starburst toward its northern region \citep{downes98,carilli00}.  The HCN/HCO$^+$ intensity ratios of Mrk\,231 and Mrk\,273 are larger than unity, whereas HCN/HCO$^+$ $<$ 1 in other galaxies.   Our results are consistent with \citet{gc08}, and the intensity ratios of HCN/HCO$^+$ and C$_2$H/HCN are all consistent with \citet{costagliola11}. At $> 1\sigma$ level, we might tentatively detect HNCO(4-3) in Mrk\,231.


\subsection{NGC\,4258}
NGC\,4258 (M\,106) is an Sab type LINER/Seyfert galaxy. There is a nuclear jet related with the anomalous arm in NGC\,4258 \citep{van der kruit 72}. Besides HCO$^+$ and HCN, the emission of C$_2$H is also strong. Other species such as H$^{13}$CO$^+$ and HCO, were also tentatively detected.
It is intriguing that the profile of C$_2$H lines of NGC\,4258 is different from other galaxies. For NGC\,4258, the emission of the hyperfine components differs only slightly in their peak intensities, whereas for other galaxies the emission is evidently dominated by the strongest component (3/2-1/2 F=2-1). This may imply that the nuclear region of NGC\,4258 covered by our beam should be optical thick, because a optical-thin case requires a 2:1 intensity ration between the two hyperfine components, as mentioned aboved. But it should be noted that NGC\,4258 is the nearest object in our sample and our observation only covered the very central region ($\sim 1 kpc$) of this galaxy, while the spectra of more distant galaxies such as Mrk\,231 and Mrk\,273 contain the emission from entire galaxy. Such difference should be considered when we compare NGC\,4258 with those galaxies of different distance, as the central part of a galaxy is usually resided with more dense gas, more active star formation and has higher optical depth.

\section{Discussion}\label{discussion}

\subsection{C$_2$H in Galaxies}
C$_2$H, a representative hydrocarbon molecule, was initially detected in some Galactic sources \citep{tucker74}. Its relative abundance in interstellar medium was found to be the same order as that of HCN and HCO$^+$. Its dipole moment \citep[$0.8D$,][]{wilson77} is between such high density gas tracers and CO. But C$_2$H(4-3) observations toward a few Galactic molecular clouds showed that this molecule arises from relatively dense gas area, with $n(\rm H_2)\sim 10^4-10^5~\rm cm^{-3}$ \citep{watt88}. Observations also showed that C$_2$H is widespread over the inner Galactic plane \citep{liszt95}, and was detected in almost all types of molecular clouds \citep{wootten80}. The rotational transitions up to $N=11-10$ ($\sim$ 1 THz) of C$_2$H, have been measured \citep{gottlieb83,sastry81,muller00}.
The abundance of C$_2$H is high in the beginning of massive star formation, and then declines in the central region as the hot core evolves. Therefore it was suggested to be a good tracer of the initial phase of massive star formation \citep{beuther08}. However, \citet{walsh10} argued that in star forming region, C$_2$H can also trace gas not directly associated with active star formation.

The formation of C$_2$H is mainly determined by the formation and evolution of ionized hydrocarbons, $\rm C^+ \rightarrow CH^+_n$ (n=2,3,4,5). CH$^+_n$ transforms into CH$_4$, then $\rm C_2H_2^+$ and $\rm C_2H_3^+$. C$_2$H molecule is formed from the dissociated recombination of $\rm C_2H_2^+$ and $\rm C_2H_3^+$ \citep[see e.g.,][for details]{turner00, meier05, beuther08}:
\begin{equation}
\rm C_2H_2^+ + \textit{e}^- \rightarrow C_2H + H,
\end{equation}
\begin{equation}
\rm C_2H_3^+ + \textit{e}^- \rightarrow C_2H + 2H/H_2,
\end{equation}

The same chemical precursor $\rm C_2H_2^+$ is shared by C$_2$H and HC$_3$N \citep{wootten80}. Other pathways of producing C$_2$H include the photodissociation of acetylene (C$_2$H$_2$) and  CH$_2$ reacting with carbon atom:
\begin{equation}
\rm C + CH_2 \rightarrow C_2H + H
\end{equation}
On the other hand, C$_2$H molecules are mainly destroyed by reacting  with oxygen atoms:
\begin{equation}
\rm C_2H + O \rightarrow CO + CH
\end{equation}


For extragalactic sources, C$_2$H(1-0) emission was first detected in M\,82 \citep{henkel88} and then in NGC\,4945 \citep{henkel90}. Until the latest molecular line surveys towards nearby galaxies, C$_2$H was  resolved in a third extragalactic source, IC\,342, showing different morphology from other dense molecular species, such as C$^{34}$S(2-1), HC$_3$N(10-9) and N$_2$H$^+$(1-0) \citep{meier05}. C$_2$H(2-1) was reported to be the second strongest line after CS in 2mm observation of NGC\,253 \citep{martin06}.
Its abundance was also found to be of the same order of magnitude in NGC\,1068 as in NGC\,253 \citep{nakajima11}. An extensive survey of C$_2$H(1-0) in external galaxies is obtained by IRAM-30m broadband receiver EMIR \citep{costagliola11}. 
Our results provide systematic detections of C$_2$H line in a sample of active galaxies. Although C$_2$H is thought to be directly linked to massive star formation, comparing to our understanding of its behavior in Galactic molecular clouds, the global property of C$_2$H in galaxies and how it is affected by galaxy activities such as AGN and starburst are still ambiguous. Further observations of C$_2$H towards active galaxies is necessary for our understanding of its properties, which are associated with star formation and/or other processes such as XDR or stellar feedback, especially in NGC 4258 and NGC 4388 without active starburst.

\subsection{Isotopic Lines}\label{discussion-iso}
As summarized in Table \ref{tbl-isotope}, we tentatively detected isotopes of HCN and HCO$^+$ in four galaxies for the first time  (on $> 2\sigma$ level). The intensity ratios I(HCN)/I(H$^{13}$CN) and I(HCO$^+$)/I(H$^{13}$CO$^+$) of NGC\,3690, NGC\,6240, and NGC\,4258 are all about 40 -- 46. In IC\,694, H$^{13}$CO$^+$ was not detected, and I(HCN)/I(H$^{13}$CN) and I(HCN)/I(HC$^{15}$N) are only 16 and 20, respectively. Previous observations implied that in nearby galaxies,  I($^{12}$CO)/I($^{13}$CO) ratio is $\sim$ 10 and [$^{12}$C]/[$^{13}$C] $>$ 40 \citep{henkel93b,henkel10,riquelme10}. Referring to I(HCN)/I(H$^{13}$CN) $\sim$ (1--3) $\times$ I($^{12}$CO)/I($^{13}$CO) \citep{paglione97}, such result is consistent with [$^{12}$CO]/[$^{13}$CO] ratio of $\geq$ 19 found in Arp\,299 \citep{aalto91}.  However, our isotopic line intensity ratios are significantly larger than \citet{costagliola11},  this difference might result from that we did not apply Gaussian fitting to derive the intensities.

We roughly assume that the filling factor and [$^{12}$C]/[$^{13}$C] abundance ratio for HCN and HCO$^+$ are equal, and we also assume that the filling factor and excitation temperatures of the isotopes of each molecule are equal. Then the the optical depth of each  isotope is  estimated using:
\begin{equation}\label{equ-isotope}
R = \frac{1-e^{-\tau_{12}}}{1-e^{-\tau_{13}}},
\end{equation}
where R is the ratio of measured integrated intensities. The ratios I(HCN)/I(H$^{13}$CN), I(HCO$^+$)/I(H$^{13}$CO$^+$) and I(HCN)/I(HC$^{15}$N) in four sources are listed in Table \ref{tbl-isotope}. [$^{12}$C]/[$^{13}$C] ratio was suggested to be 40--50 for  nearby galaxies by \citet{henkel93a}. For comparison, here we test three [$^{12}$C]/[$^{13}$C] abundance ratios of 49, 68, and 89 to derive the optical depth of HCN(1-0) and HCO$^+$(1-0). The first abundance ratio is from investigation of nearby star forming region \citep{wang09, chin99}, the second is local interstellar medium value \citep{milam05} and the third is solar system value \citep{clayton04}. Various [$^{12}$C]/[$^{13}$C] ratios have been proposed \citep[mainly from CO observations, e.g.][]{henkel93a, langer93, lucas98, casassus05}, and we adopted these representative values for crude estimation. The two [$^{14}$N]/[$^{15}$N] abundance ratios of 100 and 270 are adopted for the same reason \citep{chin99, lucas98}.
The optical depths of each species derived from solving Equation (\ref{equ-isotope}) are listed in Table \ref{tbl-tau}.

We find that the optical depths of HCN(1-0) and HCO$^+$(1-0) are low to moderate ($\tau <$ 2) in NGC\,3690, NGC\,4258, and NGC\,6240, but high ($\tau \sim 3 - 5)$ in IC\,694, considering optical-thin case always applies for the isotopic emission. Our results are consistent with the optical depths found for some other nearby galaxies \citep{nguyen92}. Adopting [$^{12}$C]/[$^{13}$C] = 68, we obtain moderate optical depth of HCN or HCO$^+$ ($\tau \sim$ 1), which seems reasonable for these gas-rich galaxies. This may be the case, because when we use HCN, H$^{13}$CN and HC$^{15}$N together to calculate the optical depth, adopting [$^{12}$C]/[$^{13}$C] = 68 and [$^{14}$N]/[$^{15}$N] = 100, we yield similar results, $\tau \sim$ 4.1 and $\tau \sim$ 5.1, for HCN in IC\,694. We can speculate that in other galaxies without detection of isotopes, the isotopic ratios should be larger, thus in these galaxies the optical depth of HCN and/or \hcop may be low to moderate.


\begin{table*}
\centering
\begin{minipage}{160mm}
\caption{Molecular Line Luminosity and Lines Ratios$^*$}
\label{tbl-ratios}
\begin{tabular}{@{}lccccccc@{}}
\hline
  & $L'_{\rm HCO^+}$ & $L'_{\rm HCN}$ & $L'_{\rm C_2H}$ &
 $\frac{\rm HCN}{\rm HCO^+}$ & $\frac{\rm C_2H}{\rm HCN}$ & $\frac{\rm C_2H}{\rm HCO^+}$\\
 Sources & (10$^8$ K km s$^{-1}$ pc$^2$) & (10$^8$ K km s$^{-1}$ pc$^2$) & (10$^8$ K km s$^{-1}$ pc$^2$) & \\
\hline
 IC 694   & 1.27$\pm$0.05 &	0.74$\pm$0.05& 	0.49$\pm$0.06&0.58$\pm$0.05 & 0.67$\pm$0.09 & 0.39$\pm$0.05 \\
 NGC 3690 & 0.82$\pm$0.06 &	0.42$\pm$0.06& 	0.46$\pm$0.07&0.51$\pm$0.08 & 1.09$\pm$0.22 & 0.56$\pm$0.09 \\
 NGC 6240 & 9.95$\pm$0.37 &	5.91$\pm$0.37& 	2.44$\pm$0.32&0.59$\pm$0.04 & 0.41$\pm$0.08 & 0.25$\pm$0.03 \\
 Mrk 231  & 7.31$\pm$0.88 &	10.3$\pm$0.88& 	2.52$\pm$1.02&1.41$\pm$0.21 & 0.24$\pm$0.10 & 0.34$\pm$0.15 \\
 Mrk 273  & 4.71$\pm$1.96 &	5.92$\pm$1.96& 	      $<$1.17&1.26$\pm$0.67 & $<$ 0.20      & $<$ 0.25      \\
 NGC 4258 &0.041$\pm$0.002&   0.023$\pm$0.002& 0.013$\pm$0.002&0.56$\pm$0.06 & 0.58$\pm$0.11 & 0.33$\pm$0.06 \\
 NGC 4388 & 0.20$\pm$0.05 &	0.08$\pm$0.05&     0.12$\pm$0.06&0.43$\pm$0.26 & 1.39$\pm$1.05 & 0.59$\pm$0.33 \\
 NGC 4593 & 0.07$\pm$0.08 &	0.12$\pm$0.08&          $<$ 0.09&1.84$\pm$2.49 & $<$ 0.68      & $<$ 1.25      \\
 NGC 5506 & 0.08$\pm$0.04 &	0.04$\pm$0.04&          $<$ 0.04&0.47$\pm$0.53 & $<$ 1.08      & $<$ 0.50      \\
 NGC 5548 & 0.09$\pm$0.10 &	 $<$ 0.05    &          $<$ 0.07& $<$ 0.52     & $\sim$1.51    & $<$ 0.78   \\
\hline
\end{tabular}\\
\medskip
$^*$Molecular line luminosity based on $L' = \pi/(4{\rm ln}2) \theta_{\rm mb}^2 I_{\rm HCN} d_L^2 (1+z)^{-3}$ \citep{gao04a}.
 Upper limits are $<$ 2 $\sigma$.
\end{minipage}
\end{table*}
 
\begin{table*}
\centering
\begin{minipage}{125mm}
\caption{Molecular isotopes$^*$}
\label{tbl-isotope}
\begin{tabular}{@{}lcccccc@{}}
\hline
 Source &
 H$^{13}$CO$^+$ & $\frac{\rm HCO^+}{\rm H^{13}CO^+}$ &
 H$^{13}$CN & $\frac{\rm HCN}{\rm H^{13}CN}$ &
 HC$^{15}$N & $\frac{\rm HCN}{\rm HC^{15}N}$ \\
\hline
 IC\,694	  &      ...            &     ...     & 0.12\phn$\pm$0.05\phn  &     16.8$\pm$9.0\phn  & 0.10$\pm$0.05 & 20.0$\pm$13\phn\phd \\
 NGC\,3690 &  0.053$\pm$0.05\phn & 40.2$\pm$43 & 	    ...        &        ...     &     ...      &     ...    \\
 NGC\,6240 &             ...     &     ...     & 0.073$\pm$0.05\phn     & 40.6$\pm$32\phn\phd   &       ...    &  ...      \\
 NGC\,4258 &  0.052$\pm$0.05\phn & 46.3$\pm$48 & 	 ...           &   ...         &        ...     &    ...      \\
\hline
\end{tabular}\\

\medskip
$^*$Molecuar isotopic line emission and line ratios of four galaxies with tentative isotope detections ($> 2\sigma$).
\end{minipage}
\end{table*}

\begin{table}
\centering
\caption{Derived optical depth$^*$}
\label{tbl-tau}
\begin{tabular}{@{}lcccccc@{}}
\hline
 & \multicolumn{3}{c}{$^{12}$C/$^{13}$C} \\ 
\cline{2-4} \cline{6-7} 
 & 49 & 68 & 89 & \\ 
\hline
NGC\,3690: \\
~~~~$\tau(\rm HCO^+)$ &  0.42\phn  &  1.2\phn  &  1.9\phn  \\
~~~~$\tau(\rm H^{13}CO^+)$ &  0.009     &  0.02     &  0.02     \\
NGC\,4258: \\
~~~~$\tau(\rm HCO^+)$ &  0.12\phn  &  0.8\phn  & 1.5\phn   \\
~~~~$\tau(\rm H^{13}CO^+)$ &  0.002     &  0.01     & 0.02      \\
NGC\,6240: \\
~~~~$\tau(\rm HCN)$ &  0.40\phn  &  1.2\phn  & 1.9\phn     \\
~~~~$\tau(\rm H^{13}CN)$ &  0.008     &  0.02     & 0.02        \\
IC\,694:  \\
~~~~$\tau(\rm HCN)$ &   2.8\phn  &  4.1\phn  &  5.4\phn  \\
~~~~$\tau(\rm H^{13}CN)$ &  0.06      &   0.06    & 0.06      \\
\hline
 & \multicolumn{2}{c}{$^{14}$N/$^{15}$N} \\
\cline{2-3} &
  100 & 270 & \\
\hline
IC\,694:  \\
~~~$\tau(\rm HCN)$~~~ &  5.1\phn  & 13.8    \\
~~~$\tau(\rm HC^{15}N)$~~~ &  0.05     & 0.05    \\
\hline
\end{tabular}\\

\medskip
$^*$Using various [$^{12}$C]/[$^{13}$C] (49, 68 and 89) and  [$^{14}$N]/[$^{15}$N] abundance ratios (100 and 270), we estimated optical depth for HCN, H$^{13}$CN, HCO$^+$, H$^{13}$CO$^+$ in four galaxies. HCN/HC$^{15}$N optical depth is also derived in IC\,694 {\it (See text for details)}.

\end{table}

\subsection{Line ratios -- dense gas tracer}
Because our observations only collected the emission from the inner region of the galaxies matching the 29$''$ beam size,  before we try to inspect the relation between the infrared luminosity and the emission line ratios, we use images of {\it Spitzer} MIPS 70$\mu m$ to determine the infrared luminosity ratio inside the 29$''$ aperture of each galaxies to match the corresponding region of molecular line emission. Although the resolution of MIPS 70$\mu m$ is lower than that of MIPS 24$\mu m$, the 24$\mu m$ emission of a galaxy may be contaminated by hot dust emission heated by obscured AGN, thus it may not be a good proxy of SFR in this study. And recent Herschel studies suggest that the 70$\mu m$ luminosity is a good tracer of the global SFR even in high-redshift galaxies \citep{magnelli11, elbaz11}.  For the most distant sources (e.g., Mrk\,231 and Mrk\,273), we skip such procedure since they were fully covered by the beam size.

\begin{figure}
\includegraphics[scale=.7]{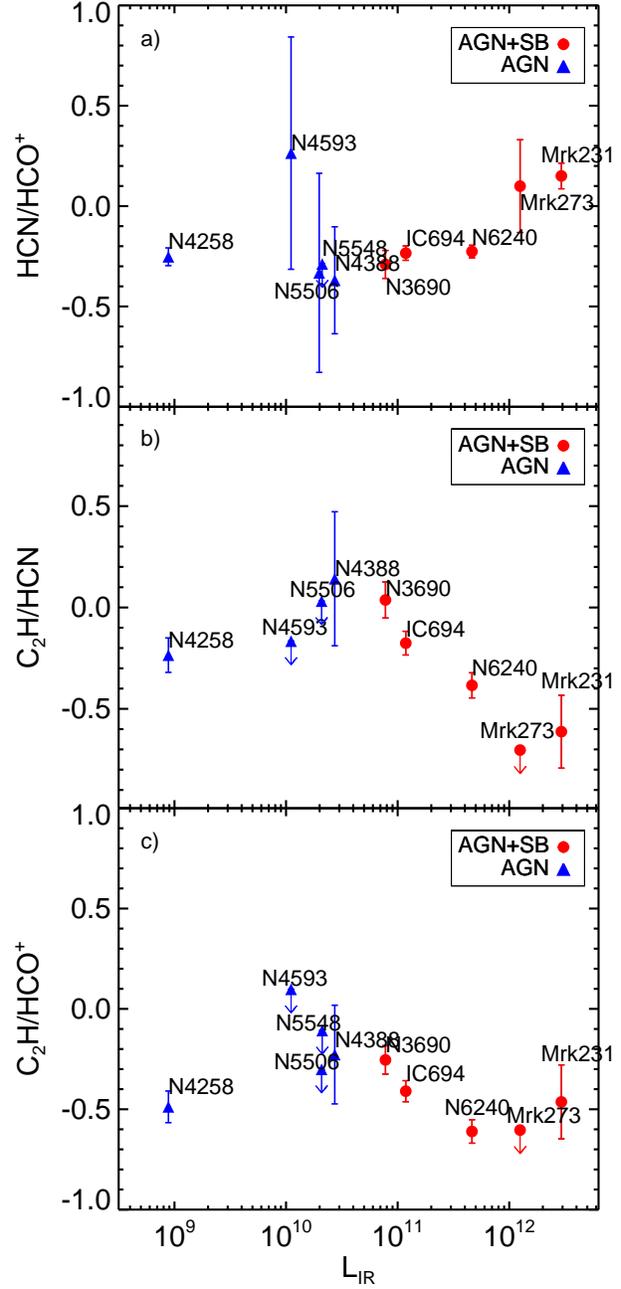} \\
 \caption{ I(HCN)/I(HCO$^+$), I(C$_2$H)/I(HCN) and I(C$_2$H)/I(HCO$^+$) plotted as a funtion of \LIR~ for both AGN+SB composite galaxies{\it (Red solid circles)} and pure AGN{\it (blue solid triangle)}. The ratios are in logarithmic scale. The infrared luminosities \LIR~ are corrected for the beam effect(see text for details).}\label{fig-ratios}
\end{figure}

\begin{figure}
\includegraphics[scale=.7]{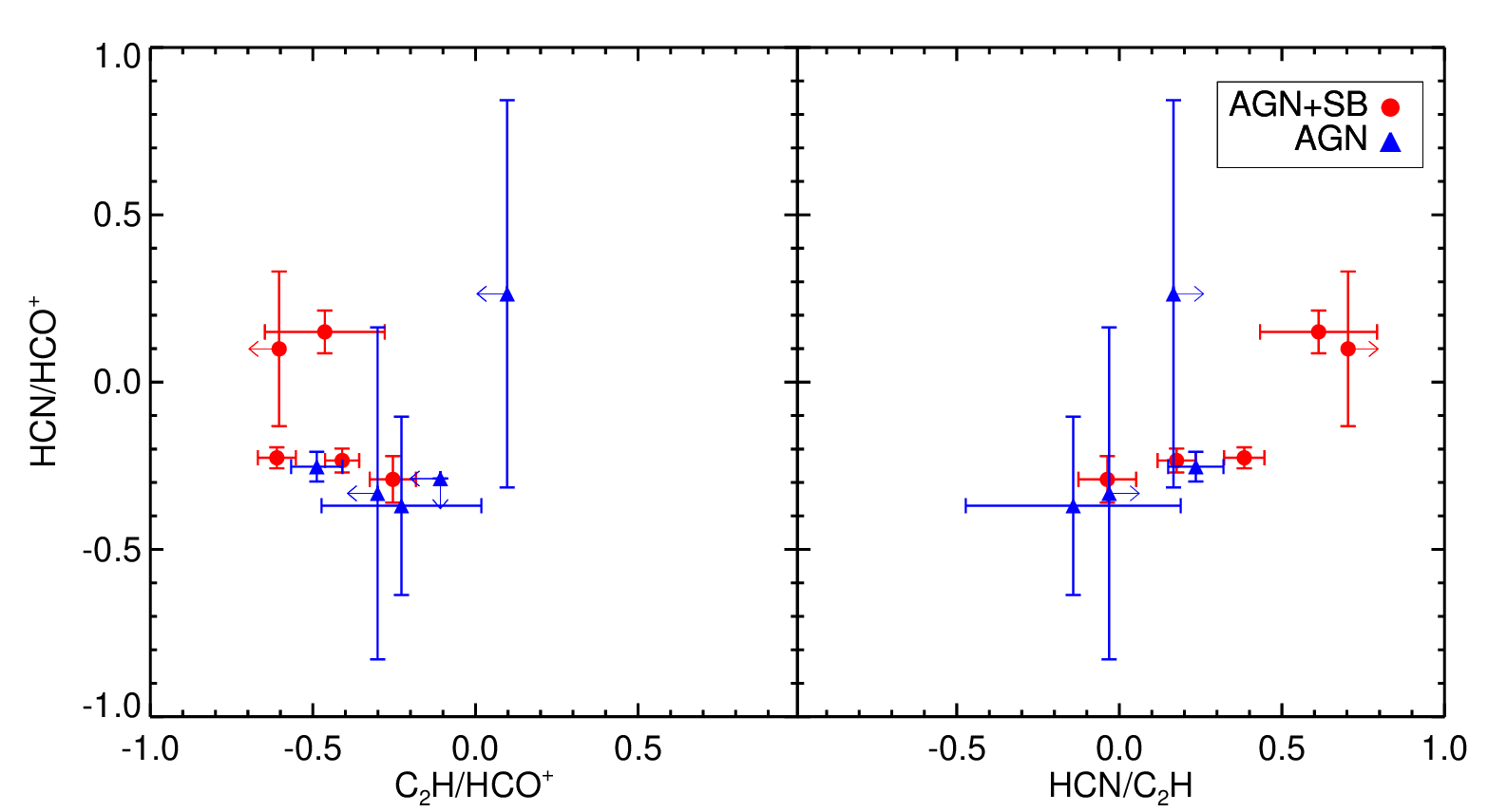} \\
\caption{{\it left panel:} I(HCN)/I(HCO$^+$) vs. I(C$_2$H)/I(HCO$^+$); {\it right panel:} I(HCN)/I(HCO$^+$) versus I(HCN)/I(C$_2$H).}
\label{fig-c2h}
\end{figure}

Ratios of HCN, HCO$^+$, and C$_2$H are listed in Table \ref{tbl-ratios}. These ratios are plotted as a function of beam-corrected \LIR\ in Figure~\ref{fig-ratios}. Intensity ratios HCN/HCO$^+$ versus C$_2$H/HCO$^+$, and HCN/HCO$^+$ versus C$_2$H/HCN are also shown in Figure \ref{fig-c2h}.

Despite our small sample and low S/N of some  spectra, the HCN/HCO$^+$ ratios span a wide range in LIRGs/ULIRGs, which are also AGN+SB composite galaxies. However, in other galaxies with $\LIR<10^{12}~\Lsun$ these ratios remain nearly constant at $\sim$ 0.5.
Such diagram resembles that of \citet{gc06, gc08}. Thereafter, HCN has been doubted for its reliability as dense gas tracer, as AGN activity may enhance HCN in the XDR environment, while keeping HCO$^+$ unaffected, resulting in a higher HCN/HCO$^+$ ratio in galaxies with prominent AGN activity \citep{gc06}. On the other hand, as a molecular ion, HCO$^+$ is easily destroyed by recombination with electrons, and it is demonstrated that this may cause HCN/HCO$^+$ to be higher in LIRGs \citep{papadopoulos07}. Furthermore, the wide excitation range among galaxies renders that simply adopting low $J$ transition of HCN and/or HCO$^+$ lines are unreliable for determining the properties of galaxies. Moreover, the critical densities ($n_{\rm crit}$) of HCN and HCO$^+$ (of the same $J$-level) differ by about one order of magnitude \citep{evans99}, thus they may trace different region in the same galaxy. This should be taken into account when one compares these molecules as dense gas mass tracers and determines the influence of galaxy activity on these species. Comprehensive multi-transition multi-species observation is needed for such kind of study \citep{greve09}.



The C$_2$H/HCN and C$_2$H/HCO$^+$ ratios plotted versus \LIR\ are also shown in panel b) and c) of Figure \ref{fig-ratios}. Since we have obtained the integrated intensities of C$_2$H simultaneously, these three molecules can be used to establish reliable different line ratios together to study the gaseous properties of the galaxies, avoiding both pointing errors and  beamsize confusions. Panel b) shows a very intriguing relation that C$_2$H/HCN seems to decrease with increasing \LIR, while in panel c) such trend is looser between C$_2$H/HCO$^+$ and \LIR. We note that  C$_2$H/HCN and C$_2$H/HCO$^+$ ratios of NGC\,4258 resemble those of the most luminous galaxies in our sample, although its \LIR\ is the lowest. This may be due to the beamsize effect that our observation only obtained the emission from the nuclear region of NGC\,4258 as it is the nearest object, while for other distant galaxies we obtained more extended gas emission from both the nuclear region and the disk. Hence it may imply that the overall molecular gas activity of the IR luminous galaxies is more extreme, which is similar to the nuclei of nearby normal galaxies.

The difference between critical density of these three molecules can explain such trend in Figure \ref{fig-ratios}. The critical densities ($n_{crit} \propto \mu^2\nu^3$) of HCN and HCO$^+$ 
are much higher than that of C$_2$H ($\mu \sim 2.98$D for HCN, $\mu \sim 3.92$D for HCO$^+$ and $\mu \sim 0.8$D for C$_2$H, respectively), so basically they trace regions of different density. First, the relatively high density gas component can be traced by HCN and HCO$^+$, and the intermediate density component can be traced by C$_2$H, thus the C$_2$H/HCN and C$_2$H/HCO$^+$ ratios will decrease toward higher \LIR~ if galaxies with higher \LIR~ contain more dense gas. Second, considering HCN traces denser gas component than HCO$^+$ does, the slope of C$_2$H/HCN versus \LIR~ will be steeper than that of C$_2$H/HCO$^+$ versus \LIR, as indicated in Figure \ref{fig-ratios}.

Following the above scenario, Figure \ref{fig-c2h} shows HCN/HCO$^+$ plotted as a function of C$_2$H/HCO$^+$ (panel a) and C$_2$H/HCN (panel b). In the previous work, ratios between HCN(1-0)/HCO$^+$(1-0) and HCN(1-0)/CO(1-0) were used to distinguish AGN- and starburst-dominated galaxies \citep{kohno05, imanishi09}. Here we attempt to replace CO(1-0) with C$_2$H(1-0) in such diagnostic diagram, as C$_2$H(1-0) reflects medium dense gas content which may also be star-forming gas as well. Due to our small sample size as well as the large uncertainties for the pure AGN galaxies, we could not separate the galaxies dominated by AGN or starburst clearly based on their location in these plots. If this is not just due to the large uncertainties, it may imply that the molecular properties are not distinct in galaxies with different activities, but mainly depend on the overall molecular gas density of the galaxy.

\begin{figure}
\includegraphics[scale=.55]{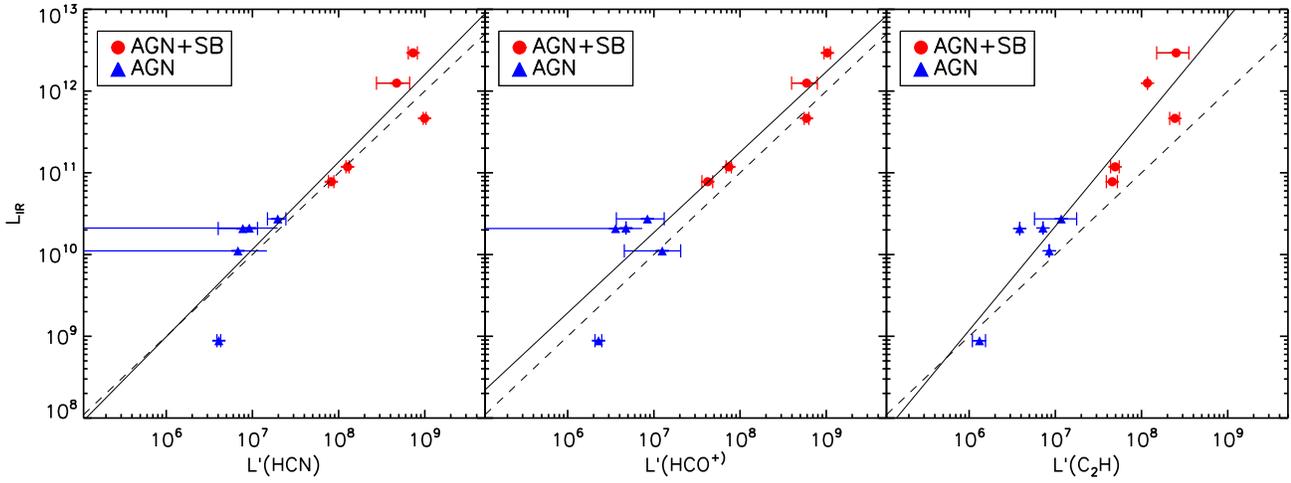}
\caption{\LIR~ as a function of molecular line luminosity $L'_{\rm HCN}$ ({\it left}), $L'_{\rm HCO^+}$ ({\it centre}) and $L'_{\rm C_2H}$ ({\it right}) ($L'$ is in K km s$^{-1}$ pc$^2$). The solid line is linear fit of the data points and dashed line is reference line with slope equals unit.}
\label{fig-l-ratio}
\end{figure}

In Figure. \ref{fig-l-ratio}, we explore the relation between molecular line luminosity and star formation rate (SFR) indicated by \LIR ~corrected for beam effect. Despite the small size of our sample, we find that the data points in the three plots show different slopes. The slope of our best linear fit to HCN (or HCO$^+$) line luminosity (log $L'$) and log \LIR~ is close to unity and  consistent to previous results \citep{gao04a, gc08}. But in log $L_{\rm C_2H}$ - log \LIR~ plot, the slope is superlinear, $\sim$ 1.3. This is consistent with the scenario that the correlation between star formation rate and gas surface or volume density, depends on the critical density of the molecular line chosen to trace the gas \citep{krumholz07, bussmann08}. Thus the different slopes may just reflect the different critical densities of HCN, HCO$^+$ and C$_2$H.

\section{Summary}
We presented broad-band observations in 3 mm band using the IRAM 30m telescope for ten active galaxies spanning three orders of magnitude in infrared luminosity. The observation had benefit of the improvement of the wide-band spectrometer onboard the IRAM 30m telescope. Our main results include:

\begin{enumerate}

\item We obtain systematic C$_2$H(1-0) emission measurements in addition to HCN(1-0) and HCO$^+$(1-0) with simultaneous observation for a sample of 10 nearby active galaxies, half of which also show starburst activity. It is an important complementarity of recent multi-species investigation of galaxies, because the chemical behavior of C$_2$H is suggested to be directly related to star forming gas.

\item We obtain tentative detections of H$^{13}$CN, HC$^{15}$N, and H$^{13}$CO$^+$ in up to four galaxies, NGC\,3690, NGC\,4258, NGC\,6240, and IC\,694. We discuss the abundance ratio and optical depth of $^{12}$C and $^{13}$C, and find that the optical depths of HCN and/or HCO$^+$ are low to moderate in NGC\,3690, NGC\,4258 and NGC\,6240 but high in IC\,694, considering all the sources are optically thin for the isotopic emission.

\item We find that HCN(1-0)/HCO$^+$(1-0) intensity ratio spans a wide range in the ultra-luminous infrared galaxies. On the other hand, C$_2$H(1-0)/HCN(1-0) and C$_2$H(1-0)/HCO$^+$(1-0) ratios in (U)LIRGs seem inversely proportional to \LIR. We speculate that these results reflect the different critical densities of these three molecules. 

\item The luminosity of C$_2$H(1-0) shows a correlation with \LIR~ in log-log scale with a slope $\sim$ 1.3, while for HCN(1-0) and HCO$^+$(1-0) their luminosities are both nearly proportional to \LIR. 

\end{enumerate}

\section*{Acknowledgments}
The authors are very grateful to the anonymous referee
for her/ his careful reading and constructive comments which 
improved the paper significantly. 
We thank the staff at the IRAM 30m telescope 
for their kind help and support during our observations.
We appreciate for the kindly help from Zhiyu Zhang with the data reduction 
and discussion.
We would also like to acknowledge the financial
support from the Natural Science Foundation of China under grants
10878010, 10633040, 10833006 and 10803002, and  the National Basic Research Program (973
program No. 2007CB815405). This project is supported by the Fundamental 
Research Funds for the Central Universities. This research has made use of NASA's
Astrophysics Data System, and the NASA/IPAC
Extragalactic Database (NED), which is operated by the Jet
Propulsion Laboratory, California Institute of Technology, under
contract with the National Aeronautics and Space Administration.
This work is based on observations made with the Spitzer Space
Telescope, which is operated by the Jet Propulsion Laboratory,
California Institute of Technology, under NASA contract 1407.

\bsp
\label{lastpage}

\end{document}